\documentclass[aps,showpacs,preprint]{revtex4}%
\usepackage{amsfonts}
\usepackage{amsmath}
\usepackage{amssymb}
\usepackage{graphicx}%
\begin{document}
\title{Adsorption and dissociation of O$_{2}$ at Be(0001): First-principles
prediction of an energy barrier on the adiabatic potential energy surface }
\author{Ping Zhang, Bo Sun, and Yu Yang}
\affiliation{LCP, Institute of Applied Physics and Computational Mathematics, P.O. Box
8009, Beijing 100088, People's Republic of China}

\begin{abstract}
The adsorption and dissociation of O$_{2}$ molecules at the Be(0001) surface
is studied by using density-functional theory within the generalized gradient
approximation and a supercell approach. The physi- and chemisorbed molecular
precursor states are identified to be along the parallel and vertical
channels, respectively. It is shown that the HH-Z (see the text for
definition) channel is the most stable channel for the molecular chemisorption
of O$_{2}$. The electronic and magnetic properties of this most stable
chemisorbed molecular state are studied, which shows that the electrons
transfer forth and back between the spin-resolved antibonding $\pi^{\ast}$
molecular orbitals and the surface Be $sp$ states. A distinct covalent weight
in the molecule-metal bond is also shown. The dissociation of O$_{2}$ is
determined by calculating the adiabatic potential energy surfaces, wherein the
T-Y channel is found to be the most stable and favorable for the dissociative
adsorption of O$_{2}$. Remarkably, we predict that unlike the other simple
$sp$ metal surfaces such as Al(111) and Mg(0001), the \textit{adiabatic}
dissociation process of O$_{2}$ at Be(0001) is an activated type with a
sizeable energy barrier.

\end{abstract}
\maketitle

\section{INTRODUCTION}

It is of great scientific importance to understand the behaviors of diatomic
molecules at solid surfaces, including their adsorption and dissociation, and
corresponding energy barriers during the bond breaking and bond formation at
the surfaces \cite{Darling1995,King1988}. Of all prototypes, the interaction
of O$_{2}$ molecules with metal surfaces has gained lots of interest for many
technologically relevant processes such as heterogeneous catalysis and
corrosion \cite{Kung}. Theoretically, \textit{ab initio} modeling has been
successfully used over a wide range to study the adsorption and dissociation
of O$_{2}$ molecules at transition metal surfaces. By calculating the
adiabatic potential energy surface (PES), it has been found that O$_{2}$
molecules will spontaneously dissociate while adsorbing at reactive transition
metal surfaces like iron (Fe) \cite{Blonski}. For noble transition metals like
gold (Au) \cite{Yotsuhashi}, silver (Ag) \cite{Nakatsuji,Gra1996}, platinum
(Pt) \cite{Yotsuhashi,Eichler1997,Eichler} and Nickel (Ni) \cite{Eichler}, the
adsorption of O$_{2}$ turns out to depend on the ambient temperature.
Remarkably, in all above transition metal systems, the concept of adiabatic
PES works very well in explaining and predicting a large amount of
physical/chemical phenomena during dissociation process of O$_{2}$.

When the attention is focused on the simple $sp$ metals such as Al(111),
however, an uncomfortable gap opens between \textit{ab initio} prediction and
experimental observation. The most notable is the long-term enigma of low
initial sticking probability of thermal O$_{2}$ molecules at Al(111), which
has been measured by many independent experiments \cite{Ertl,Osterlund} but
cannot be reproduced by adiabatic state-of-the-art density functional theory
(DFT) calculations \cite{Honkala,Your2001,Your2002}. The central problem is
that the adiabatic DFT calculations were unable to find any sizeable barriers
on the adiabatic PES, whose presence, however, is essential for explanation of
the experimental finding. This has led to speculations that nonadiabatic
effects may play an important role in the oxygen dissociation process at the
Al(111) surface
\cite{Your2001,Kas1974,Kas1979,Kat2004,Wod2004,Hell2003,Hell2005}. Recently, a
semiquantitative agreement with the experimental data is achieved by
nonadiabatically confining the trajectories of the approaching O$_{2}$
molecules to the spin-triplet PES \cite{Beh2005}. Until now, however, it still
remains unclear how and in what chemical circumstance this spin-selection rule
is reasonable, and different opinions exist in literature \cite{Fan2006}. For
oxygen dissociation at another simple metal, Mg(0001), first-principles DFT
calculation also shows the lack of a barrier on the adiabatic PES
\cite{Hellman2005}, which is in strong disagreement with experimental
observation \cite{Dri1999,Aba2004} of a low sticking coefficient of O$_{2}$ at
Mg(0001). This discrepancy was ascribed, in a similar manner with that in the
O$_{2}$/Al(111) system, to nonadiabaticity in the dissociation process of
oxygen molecules \cite{Hellman2005}.

Therefore, it becomes clear that the theoretical study of oxygen dissociation
at the simple $sp$ metals is still far from its maturity. In particular,
considering the fact that up to now most of the theoretical results and
conclusions are only based on the O$_{2}$/Al(111) and O$_{2}$/Mg(0001)
prototypes, workers are thus confronted with an important question: Does it
retain necessary or valid for all simple $sp$ metals (at least, for metals
with similar elemental valence electrons to Al or Mg) to take into account
such nonadiabaticity as spin-selection rule in obtaining an activated PES?
Motivated by this question, in this paper we have carried out a systematic
\textit{ab initio} investigation of the adsorption and dissociation of oxygen
molecules at Be(0001) surface. Beryllium has the same crystal structure and
valence electrons as magnesium does. Subsequently, it is attempting for one to
derive that the behaviors of oxygen dissociation at Be(0001) are the same as
or similar to that at Mg(0001). Our results, however, show that this is not
true. The most distinct is that in the present O$_{2}$/Be(0001) system, our
calculated adiabatic PES displays sizeable energy barriers along the
dissociation paths. This partially but definitely answers the above question.
That is, in the DFT calculations of the simple-metal systems, the inclusion of
nonadiabatic effects is not always indispensable for the presence of molecular
dissociation barrier.

Besides this basic point of interest, our present study is also motivated by
the fact that Be has vast technological applications due to its high melting
point and low weight. During these applications, surface oxidation as the main
kind of corrosion always needs to be prevented. Thus a systematic study on the
adsorption and dissociation of O$_{2}$ molecules at Be surfaces, i.e., the
initial stage during the surface oxidation process, should be done. Moreover,
Be is also a getter in experimental nuclear fusion reactors to adsorb residual
gases such as O$_{2}$ and H$_{2}$O in the plasma vessel, improving the plasma
cleanliness \cite{Zalkind}, which also makes it highly meaningful to study the
adsorption and dissociation of O$_{2}$ at Be surfaces. At present there are
only very few experimental data on the oxidation of Be, reporting that the
surface oxidation begins at an oxidation nucleation center, followed by a
spreading growth from the center \cite{Zalkind,O-Be-2,O-Be-3,O-Be-4}, while
\textit{ab initio} studies are entirely lacking. This as a whole encourages us
to theoretically report a systematic investigation on the O$_{2}$/Be(0001)
system. The rest of this paper is organized as follows: In Sec. II the
computational method is briefly described. In Sec. III results for the free
oxygen molecule, the bulk Be and clean Be(0001) surface are given. In Sec. IV
and Sec. V we present our results for O$_{2}$ adsorption and dissociation at
Be(0001) surface, respectively. Finally, we close our paper with a summary of
our main results.

\section{COMPUTATIONAL METHOD}

The density-functional theory (DFT) total energy calculations were carried out
using the Vienna \textit{ab initio} simulation package \cite{Vasp} with the
projector-augmented-wave (PAW) pseudopotentials \cite{Paw} and plane waves
\cite{Plane Wave}. The so-called \textquotedblleft repeated
slab\textquotedblright\ geometries were employed \cite{Slab}. This scheme
consists in the construction of a unit cell of an arbitrarily fixed number of
atomic layers identical to that of the bulk in the plane of the surface
(defining the dimensional cell), but symmetrically terminated by an
arbitrarily fixed number of empty layers (the \textquotedblleft\textit{vacuum}%
\textquotedblright) along the direction perpendicular to the surface. In the
present study, the clean $p(2\times2)$-Be(0001) surface was modeled by
periodic slabs consisting of nine Be layers separated by a vacuum of 20 \AA ,
which was found to be sufficiently convergent. Oxygen molecules were
symmetrically introduced on both sides of the slab. During our calculations,
the positions of the outmost three Be layers as well as the O$_{2}$ molecules
were allowed to relax until the forces on the ions were less than 0.02
eV/\AA , while the central three layers of the slab were fixed in their
calculated bulk positions. The plane-wave energy cutoff was set 400 eV. After
a careful convergence analysis, we used a $11\times11\times1$ $k$-point grid
for the $p(2\times2)$ cell with Monkhorst-Pack scheme \cite{Pack}.
Furthermore, the generalized gradient approximation (GGA) of Perdew \textit{et
al}. \cite{GGA-2} for the exchange-correlation potential was employed since
the GGA results has been previously validated for the bulk Be
\cite{BulkLattice}. A Fermi broadening \cite{Fermi Broaden} of 0.1 eV was
chosen to smear the occupation of the bands around $E_{F}$ by a finite-$T$
Fermi function and extrapolating to $T\mathtt{=}0$ K.

\section{BULK Be, CLEAN Be(0001) SURFACE, AND FREE OXYGEN MOLECULE}

\begin{figure}[ptb]
\begin{center}
\includegraphics[width=0.5\linewidth]{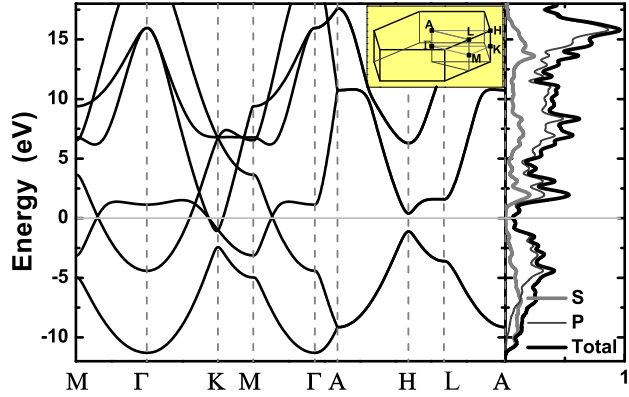}
\end{center}
\caption{(Color online). Band structure (left panel) and orbital-resolved DOS
(right panel) of the bulk hcp Be. The Fermi energy is set at zero. The inset
shows the Brillouin zone. }%
\end{figure}

First, the total energy of the bulk hcp Be was calculated to obtain the bulk
lattice parameters. The calculated lattice constants are 2.26 and 3.56 \AA ~
respectively for $a$ and $c$, according well with experimental values
\cite{BulkLattice2} of 2.285 and 3.585 \AA . The band structure and
orbital-resolved density of states (DOS) of bulk Be is shown in Fig. 1.
Typically, the DOS of bulk Be has a relatively small value at the Fermi energy
($E_{F}$), because there exists a Dirac point at the Fermi energy in the band
structure along the $\mathbf{\Gamma}\mathtt{-}\mathbf{M}$ direction, and a
wide band gap along the $\mathbf{\Gamma}\mathtt{-}\mathbf{A}$ direction. This
anisotropic property of Be differs from that of other alkali-earth metals,
whose band structures around the Fermi energy are always nearly free-electrons
like. Although the electronic configuration of elemental Be is 1$s^{2}$%
2$s^{2}$, one can see from Fig. 1 that the 2$p$ states play an important role
in the DOS behaviors below $E_{F}$, indicating a strong hybridization between
$s$ and $p$ electronic states in bulk Be.

The band structure and total DOS for the clean Be(0001) film are shown in Fig.
2. Compared to the bulk case (Fig. 1), one can see from Fig. 2 that there
exists no longer Dirac points or band gaps along any direction. Moreover, the
DOS at E$_{F}$ is prominently enhanced. This result accords well with previous
studies \cite{Surface-state,Surface-state-1}. Similar results have also been
observed for W (110) and Mo (110) films \cite{Surface-state-2}. From the inset
in Fig. 2, we can see that the surface electronic states around $E_{F}$ mainly
accumulate within the two topmost Be layers. Further detailed wavefunction
analysis shows that these states are mainly Be $2p$ states. Our clean-surface
calculation shows that due to this pronounced surface charge redistribution,
the two outmost Be(0001) layers relax significantly from the bulk values. The
first-second interlayer contraction is 3.8\% and the second-third interlayer
expansion is nearly 1.2\%, which is in agreement with recent first-principles
calculations \cite{Thermal-1} and comparable with experimental measurements
\cite{Relax-2}.

\begin{figure}[ptb]
\begin{center}
\includegraphics[width=0.5\linewidth]{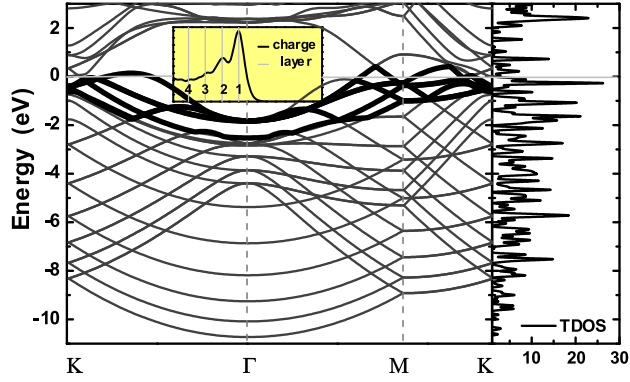}
\end{center}
\caption{(Color online). Band structure (left panel) and total DOS (right
panel) of the clean $p$($2\times2$) Be(0001) surface with nine atomic layers
included in the supercell. The Fermi energy is set at zero. The inset shows
the vertival distribution of the charge density close to the surface (Nos. 1-4
enumerate the atomic layers from the surface). }%
\end{figure}

The total energies of the isolated O atom and free O$_{2}$ molecule are
calculated in an orthorhombic cell of scale $13\times11\times17$ \AA $^{3}$~
with a $(3\times3\times3)$ $k$-point mesh for the Brillouin zone sampling. The
spin-polarization correction has been included. The binding energy of O$_{2}%
$\ is calculated to be $1/2E_{b}^{\mathbf{O}_{2}}$=2.89 eV per atom and the
O-O bond length is about 1.235 \AA . These results are typical for
well-converged DFT-GGA calculations. Compared to the experimental
\cite{Molecule Expt} values of 2.56 eV and 1.21 \AA ~ for O binding energy and
bonding length, the usual DFT-GGA result always introduces an overestimation,
which reflects the theoretical deficiency for describing the local orbitals of
the oxygen.%

\begin{figure}[tbp]
\begin{center}
\includegraphics[width=0.4\linewidth]{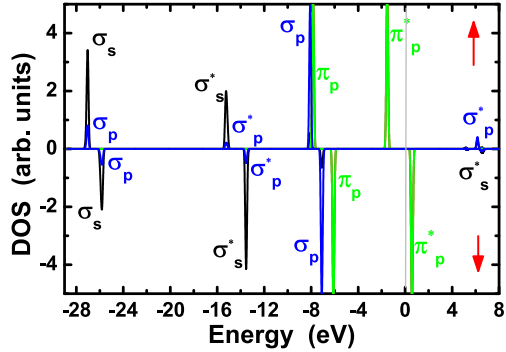}
\end{center}
\caption
{(Color online). The calculated spin-polarized density of states for the molecular orbits (MO's) of the free O$_{2}%
$ molecule. The Fermi energy level is set at zero.}
\end{figure}%
In order to show the general charge-transfer or redistribution effects on
O$_{2}$ molecular orbits (MO's) after adsorption, here we first calculate and
plot in Fig. 3 the MO-resolved local density of states of the free O$_{2}$
molecule. Typically, the bonding $\sigma$ MO's are lower in energy than the
bonding $\pi$ MO's, for both spins. This ordering of the MO's has important
consequences for the molecular bond to the surface. Also it reveals in Fig. 3
that spin splittings are sizeable ($\mathtt{\sim}2$ eV) for both bonding and
antibonding MO's. This reflects Hund's spin rule, which describes ground state
as spin-polarized for O$_{2}$ ($S\mathtt{=}1$). The highest occupied MO (HOMO)
and the\ lowest unoccupied MO (LUMO) are the spin-up and spin-down antibonding
$\pi^{\ast}$ MO's, respectively. Our calculated picture of the MO's for a free
O$_{2}$ molecule is in good agreement with the previous theoretical reports
\cite{TM-Pt-Pa-Ni,Al-5}.

\section{THE ADSORPTION OF O$_{2}$ MOLECULE}

There are four high-symmetry adsorption sites on the Be(0001) surface,
including top (T), hcp hollow (HH), fcc hollow (FH), and bridge (B) sites
depicted in Fig. 4(a). In this study, we construct twelve initial structures
with high symmetries by orienting the O$_{2}$ molecule at the four
high-symmetry sites respectively along the $X$ (i.e., [11$\bar{2}0$]), $Y$
(i.e., [$\bar{1}100$]) and $Z$ (i.e., [$0001$]) directions. We also construct
several low-symmetry initial structures by rotating the O$_{2}$ molecule in
the $XY$, $YZ$, and $XZ$ planes with small angles. In all the initial
configurations, the heights of the O$_{2}$ molecules are set at $h_{0}%
\mathtt{=}$4 \AA , see Fig. 3(b). It is found within our expectation that
after geometry optimization, these low-symmetry structures will either relax
into the high-symmetry ones or be less stable with a lower adsorption energy
than the high-symmetry ones. This is similar to what has been observed in
studying the adsorption of O$_{2}$ molecules at Pb(111) \cite{Yang}. After
geometry optimization, it is found that all the parallel adsorption states
with the O$_{2}$ molecule lying down on the substrate surface are stable. For
the vertical entrances with the O$_{2}$ molecule oriented perpendicular to the
surface, one exception is the B-Z entrance (namely, the O$_{2}$ molecule at
the bridge site with the O-O bond along the $Z$ direction), which turns to
evolve into the HH-Z adsorption configuration after geometry optimization.

\begin{figure}[ptb]
\begin{center}
\includegraphics[width=0.3\linewidth]{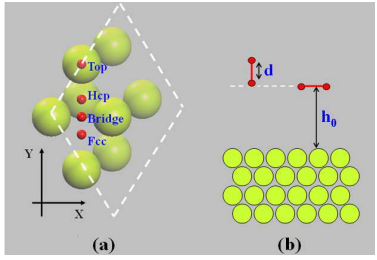}
\end{center}
\caption{(Color online) (a) The p($2\times2$) surface cell of Be(0001) and
four on-surface adsorption sites. Here only the outmost two layers of the
surface are shown. (b) The sketch map showing that the molecule (with vertical
or parallel orientation) is initially away from the surface with a hight
$h_{0}$. Note that although the atomic radius of O is much larger than that of
Be, for convenience in depicting the adsorption positions, here and in the
following we use larger circles to denote Be atoms.}%
\label{fig1}%
\end{figure}

One central quantity tailored for the present study is the average adsorption
energy of the adsorbed oxygen species defined as
\begin{equation}
E_{ad}=-\frac{1}{N_{\mathrm{O}}}[E_{\mathrm{O_{2}/Be(0001)}}%
-E_{\mathrm{Be(0001)}}-N_{\mathrm{O}}\times\frac{1}{2}E_{\mathrm{O_{2}}}],
\end{equation}
where $N_{\text{O}}$ is the total number of O atoms presented in the
supercell, $E_{\mathrm{O_{2}/Be(0001)}}$, $E_{\text{Be(0001)}}$, and
$E_{\text{O}_{2}}$ are the total energies of the slab containing oxygen, of
the corresponding clean Be(0001) slab, and of a free O$_{2}$ molecule
respectively. According to this definition, a positive value of $E_{ad}$
indicates that the adsorption is exothermic (stable) with respect to a free
O$_{2}$ molecule and a negative value indicates endothermic (unstable) reaction.

Starting from the above mentioned entrances and after geometry optimization,
the obtained molecular adsorption energy ($E_{ad}$), molecular magnetic moment
(MM), work function ($\Phi$), adsorption height ($h$), and O-O bond length
($d$) are listed in Table I. From Table I the following molecular adsorption
features are revealed: (i) For the O$_{2}$ molecule in parallel with Be(0001)
surface, the relaxed molecular structures vary very little compared to the
initial molecular structures, which suggests the molecular adsorption to be
the physisorption in these parallel channels. In these cases, the relaxed
adsorption height is around 3.9 \AA . The O-O bond length is 1.236 \AA ,
displaying a negligible expansion when compared to that of a free O$_{2}$
molecule (1.235 \AA ). The molecular MM almost saturates at its free value of
2.0 $\mu_{B}$, also suggesting the unaffected molecular orbitals in these
physisorbed structures. In addition, the work function of the adsorbed
Be(0001) surface is almost identical to that of the clean Be(0001) surface,
implying negligible charge transfer between the parallel O$_{2}$ adsorbate and
the surface Be atoms. Finally, the calculated molecular adsorption energy is
only $\mathtt{\sim}23$ meV, which is so small that a little thermal
fluctuation may result in desorption of the O$_{2}$ adsorbate with parallel
O-O bond from the Be(0001) surface. This result is consistent with the recent
experimental observation that the reactivity of oxygen atoms with Be(0001)
surface is greater than that of oxygen molecules with Be(0001) surface
\cite{O-Be-4}. On the other hand, one may question whether, when overcoming
the energy barrier by artificially setting the initial metal-molecule distance
$h_{0}$ to be lower than the above obtained physisorption height ($\sim$3.9
\AA ), new kinds of molecular adsorption state with parallel O-O bond will
occur or not. For this, a large amount of calculations with more lower initial
adsorption heights have been carried out and no other parallel molecular
physi- or chemisorption states have been found, for details see discussion in
the next section; (ii) For the O$_{2}$ adsorbate with the O-O bond oriented
perpendicular to Be(0001) surface, as shown in Table I, the relaxed molecular
structures display a signature of chemisorption. Among them the HH-Z
configuration is most stable with the largest molecular adsorption energy of
$\sim$0.5 eV (per atom). In this channel the adsorption height is decreased to
be 1.58 \AA , and prominently, the molecular MM is largely decreased to be 0.8
$\mu_{B}$, which suggests a pronounced charge redistribution among the MO's
via interaction with the Be(0001) surface. The O-O bond in this most stable
chemisorption channel is 1.471 \AA , indicating a large expansion from the
free O$_{2}$ molecule and a fundamental weakening of the molecular bonding. In
addition, the work function change is also prominent, implying an observable
charge redistribution between the chemisorbed O$_{2}$ molecule and the
Be(0001) surface. Meanwhile, during the chemisorption of the O$_{2}$ molecule,
the Be(0001) surface is also influenced. Specifically, the three Be atoms
around the adsorbed O$_{2}$ molecule of the HH-Z entrance are pulled out by
about 0.3 \AA . Recalling that the bulk BeO has an unusual wurtzite structure,
it is within one's expectation that O$_{2}$ chemisorption in the HH-Z channel
(instead of the FH-Z channel) is most energetically favorable. Our additional
calculations of the atomic oxygen adsorption at Be(0001) also show that the
hcp hollow site is most stable for atomic oxygen adsorption.

\begin{table}[ptb]
\caption{The calculated adsorption energy per atom ($E_{ad}$), molecular
magnetic moment (MM), work function ($\Phi$), adsorption height ($h$), and O-O
bond length ($d$) for physi- and chemisorptions along different channels.}%
\label{table1}
\begin{tabular}
[c]{ccccccc}\hline\hline
& Channel & $E_{ad}$ (meV) & MM ($\mu_{B}$) & $h$ (\AA ) & $d$ (\AA ) & $\Phi$
(eV)\\\hline
& T-X & 23.6 & 1.97 & 3.92 & 1.236 & 5.44\\
& T-Y & 23.1 & 1.95 & 3.91 & 1.236 & 5.44\\
& B-X & 23.9 & 1.97 & 3.91 & 1.236 & 5.44\\
Phys. & B-Y & 22.5 & 1.97 & 3.93 & 1.236 & 5.46\\
& HH-X & 23.8 & 1.99 & 3.91 & 1.236 & 5.45\\
& HH-Y & 23.3 & 2.0 & 3.9 & 1.236 & 5.43\\
& FH-X & 23.7 & 1.99 & 3.92 & 1.236 & 5.46\\
& FH-Y & 22.6 & 1.99 & 3.92 & 1.236 & 5.46\\\hline
& T-Z & 75.2 & 1.5 & 2.1 & 1.272 & 6.76\\
Chem. & HH-Z & 506.1 & 0.8 & 1.15 & 1.471 & 7.78\\
& FH-Z & 377.6 & 0.9 & 1.28 & 1.436 & 7.63\\\hline\hline
\end{tabular}
\end{table}

To clarify the bonding interaction between the chemisorbed O$_{2}$ molecule
(in the HH-Z channel) and the Be(0001) surface, we calculate and plot in Fig.
5(a) the electron-density difference $\Delta\rho(\mathbf{r)}$, which is
obtained by subtracting the electron densities of noninteracting component
systems, $\mathbf{\rho_{\mathrm{Be(0001)}}(r)+\rho_{\mathrm{O_{2}}}(r)}$, from
the density $\mathbf{\rho_{\mathrm{O_{2}/Be(0001)}}(r)}$ of the O$_{2}%
$/Be(0001) system, while retaining the atomic positions of the component
systems at the same location as in O$_{2}$/Be(0001). A positive $\Delta
\rho(\mathbf{r)}$ obviously represents charge accumulation, whereas a negative
$\Delta\rho(\mathbf{r)}$ represents charge depletion. One can see from Fig.
5(a) (where the contour spacing is 0.02$e$/\AA $^{3}$) that the charge
redistribution mainly occurs at the surface and involves the chemisorbed
O$_{2}$ molecule and the two topmost Be (0001) layers. It is apparent that
upon molecular chemisorption, the occupation of the surface and subsurface Be
$sp$ states is decreased, while the occupation of O$_{2}$ $\pi^{\ast}$
antibonding MO's is increased, which suggests an electron transfer from the
former to the latter. Meanwhile, it reveals in Fig. 5(a) that the occupation
of other bonding ($\sigma$, $\pi$) and antibonding MO's ($\sigma^{\ast}$) is
also decreased. The target states for these transfered MO charges are of
course again the O$_{2}$ $\pi^{\ast}$ antibonding MO's, as shown in Fig. 5(a).
A well-known harpooning mechanism is responsible for this inter-MO charge
transfer. Prominently, the electron redistribution of the O$_{2}$ $\pi^{\ast}$
antibonding MO's is largely asymmetric with respect to the two O atoms. The
reason is that the lower O atom in the O$_{2}$ molecule is in a complex
bonding state. Besides interacting with the upper O atom to form a
weakly-bound molecule, this lower O atom also strongly interacts with the
surface and subsurface Be atoms via a mixed ionic/covalent bonding mechanism.
In particular, it is the strong covalency in the Be-O bond that heavily
distorts the charge distribution of the O$_{2}$ $\pi^{\ast}$ antibonding MO's.
To our knowledge, this strong covalency in the bond between the metal surface
and oxygen molecule has not been theoretically reported in previous studies.
On the whole, the charge transfer to the $\pi^{\ast}$ antibonding MO's, as
well as the strong covalency in the Be-O bond, weaken the molecular bond and
spin polarization. For spin density of the chemisorbed O$_{2}$ molecule in the
HH-Z channel, see Fig. 5(b).

\begin{figure}[ptb]
\begin{center}
\includegraphics[width=0.6\linewidth]{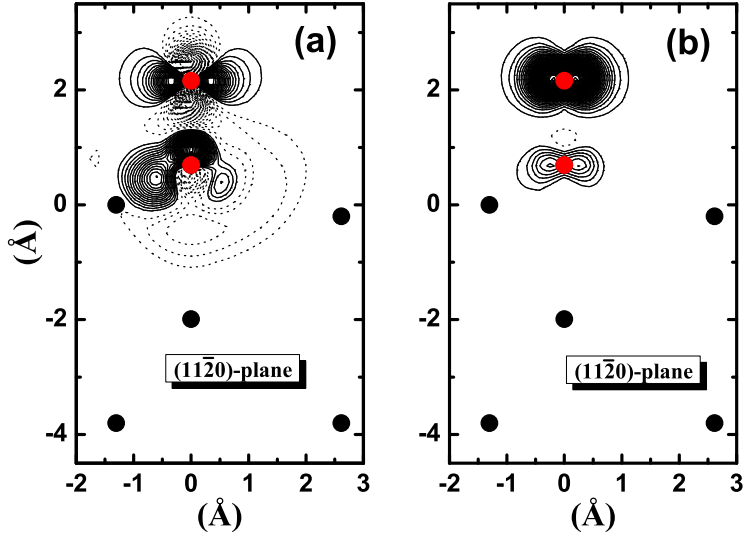}
\end{center}
\caption{(Color online) Contour plots of the (a) difference electron density
and (b) the spin density for the relaxed O$_{2}$/Be(0001) slab with O$_{2}$
molecule chemisorbed along the HH-Z entrance. Solid and dotted lines denote
accumulated and depleted densities, respectively. }%
\end{figure}

\begin{figure}[ptb]
\begin{center}
\includegraphics[width=0.5\linewidth]{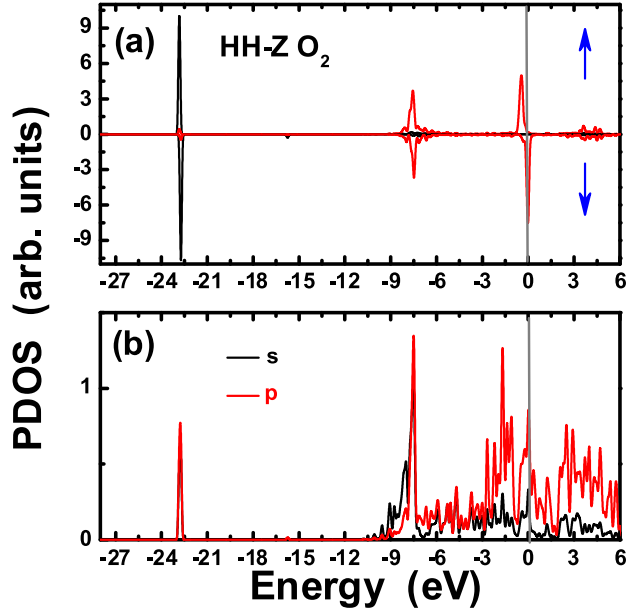}
\end{center}
\caption{(Color online) The orbital-resolved PDOS for the chemisorbed O$_{2}$
molecule along the HH-Z channel and (b) for the topmost Be layer. The Fermi
energy is set at zero. Note that for O$_{2}$ the PDOS is plotted in a
spin-split form. }%
\end{figure}

In order to gain more insights into the precise nature of the chemisorbed
molecular state in the O$_{2}$/Be(0001) system, the orbital-resolved
site-projected densities of states (PDOS) for the O$_{2}$ molecule (in the
most stable HH-Z entrance) and the topmost Be layer are plotted in Fig. 6(a)
and 6(b), respectively. By comparison with the case of a free O$_{2}$
molecule, one can see that the MO properties of the chemisorbed O$_{2}$ in the
HH-Z entrance undergo the following fundamental changes: (i) The spin-split
PDOS signature (peaks) for the two energy-lowest bonding and antibonding
$\sigma$ MO's for both spins in Fig. 3 changes to vanish in Fig. 6(a),
indicating their breaking in interacting with the Be(0001) surface. This is in
accord with the electron-density difference result shown in Fig. 5(a) that
charges mostly flow out of these two MO's; (ii) The PDOS peaks for the two
nearly-degenerate bonding $\sigma_{p}$ and $\pi_{p}$ MO's around $E$=$-$7.5 eV
are broadened upon chemisorption to merge together, with the amplitude
becoming much weaker than that in free O$_{2}$. Also, the spin splitting of
these two bonding MO's vanishes. These features consistently reveal that
although these two bonding MO's do not lose their nature as the MO's, they are
largely reshaped due to their hybridization with the $sp$ orbitals of the
surface and subsurface Be atoms. This metal-molecule hybridization is so
strong that there develops, as shown in Fig. 6, a sharp $sp$-hybrid peak
around $E$=$-$7.5 eV in the PDOS of the topmost Be layer; (iii) The spin-up
antibonding $\pi_{p}^{\ast}$ MO shifts up in energy towards $E_{F}$ and is
partially depopulated by an observable amount of charges. As a compensation,
the spin-down antibonding $\pi_{p}^{\ast}$ MO shifts down towards $E_{F}$ and
becomes partially occupied. Obviously, the spin rule prohibits direct charge
transfer from the spin-up to spin-down antibonding $\pi_{p}^{\ast}$ MO.
Therefore, the driving or intermediate factor for this charge decrease in the
spin-up and increase in the spin-down $\pi_{p}^{\ast}$ MO's is due to the
metallic Be(0001) surface, which accepts electrons from the spin-up $\pi
_{p}^{\ast}$ MO and then donates electrons to the spin-down $\pi_{p}^{\ast}$
MO. From this aspect, the present result is consistent with the well-known
harpooning mechanism, which describes adsorbate-metal interaction by a
simultaneous transfer of electrons from the adsorbing molecule into the
unoccupied metal states (direct bonding) and a back donation of electrons from
occupied metal states into the antibonding adsorbate orbitals. In the present
case, this exchange is achieved through interaction with the $sp$ orbitals of
the Be(0001) surface; (iv) Even more interestingly, to compensate the breaking
of the two energy-lowest bonding and antibonding $\sigma$ MO's, there develops
a new hybrid peak at $E$=$-$22.7 eV in the PDOS of the system. From Fig. 6 one
can see that this peak is characterized by a strong hybridization of O $2s$
and Be $sp$ states.

\section{THE DISSOCIATION OF O$_{2}$ MOLECULES}

\begin{figure}[ptb]
\begin{center}
\includegraphics[width=0.6\linewidth]{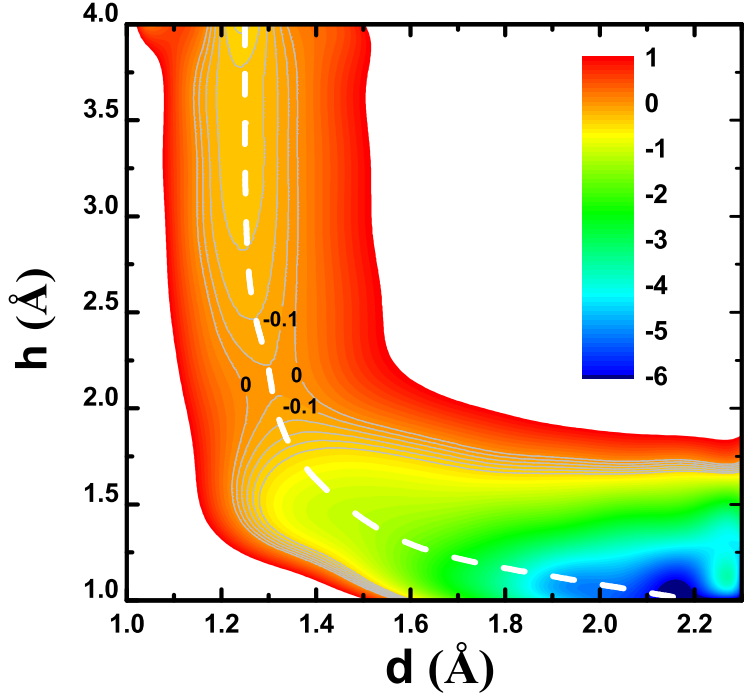}
\end{center}
\caption{(Color online) Color filled contour plot of the potential-energy
surface for O$_{2}$ dissociation at Be(0001) along the T-Y channel, as
functions of the O$_{2}$ bond length $d$ and distance $h$ (from the surface).
The contour spacing is 0.1 eV. The dashed line indicates the reaction pathway.
}%
\end{figure}

In order to deepen our understanding of the initial stage of oxidation at
Be(0001) surface from a theoretical point of view, we also calculate the
potential energy surfaces (PES's) for O$_{2}$ molecules on the Be (0001)
surface, which depicts the adiabatic dissociation path with the lowest energy
barrier. Since it has been shown that the PES given by the density functional
theory gives the result that can be compared with experiments, we expect that
our calculated PES provides the qualitative feature of the molecular
dissociation process in the present specific O$_{2}$/Be(0001) system. Figure 7
shows the obtained PES as a function of the distance $h$ from the surface and
the bond length $d$ of the O$_{2}$ molecule along the T-Y channel, which,
after a large amount of calculations, turns out to be the most favorable
dissociative adsorption channel with largest adsorption energy (4.14 eV/atom)
and lowest activation barrier (0.23 eV).

One can clearly see from Fig. 7 the process from initial molecular
physisorption with a height of 3.9 \AA ~ and bond length of 1.23 \AA ~ to the
final dissociative adsorption. The most distinct feature in Fig. 7 is that the
calculated adiabatic PES exhibits a sizeable energy barrier toward
dissociative adsorption. The transition state for this barrier is located at a
distance of 2.2 \AA ~ from the Be surface, where the O-O bond length is
elongated to 1.30 \AA . Correspondingly, if we initially put O$_{2}$ in the
T-Y channel at a height lower than 2.2 \AA , then do geometry optimization,
the O$_{2}$ molecule will spontaneously dissociate with the two O atoms moving
to the hcp hollow and fcc hollow sites, respectively. Considering that the
previous \textit{ab initio} calculations have failed to obtain a barrier on
the adiabatic PES for oxygen dissociation on the usual simple metals such as
Al(111) and Mg(0001), obviously, our present prediction of the activated PES
in the O$_{2}$/Be(0001) system is unique. In particular,\ since beryllium has
the same crystal structure and valence electrons as magnesium does, it is
attempting to assume that the behavior of oxygen dissociation at Be(0001) is
the same as or similar to that at Mg(0001). The result in Fig. 7, however,
shows that this intuitive expectation needs to be reshaped.

In the case of Al(111) or Mg(0001), the failure of adiabatic DFT calculations
in producing an activated-type PES has been ascribed to the unphysical output
that charge transfer occurs even at large molecule-metal distance, which has
led to speculations that nonadiabatic effects may play an important role in
the oxygen dissociation process at these metal surfaces with the simple $sp$
bands
\cite{Your2001,Kas1974,Kas1979,Kat2004,Wod2004,Hell2003,Hell2005,Beh2005,Hellman2005}%
. The unphysical charge transfer at large molecule-metal distance is generally
caused by the requirement in the adiabatic description that the electron
chemical potential of the O$_{2}$ molecule aligns with that of the metal
surface in the combined system (metal plus molecule). On the other hand,
however, this requirement does not necessarily results in artificial charge
transfer between the two largely-separated subsystems. Therefore, there is no
universal criterion to judge whether the adiabaticity during molecular
dissociation process breaks down or not. The molecule-metal interaction is so
species-sensitive that even for the two metals with the same crystal structure
and valence electrons, the molecular dissociation process at the surfaces can
display qualitatively different behaviors. To make it more sense, we have
carried out a comparative study on the long-distance charge transfer effect by
choosing the metal surfaces as Be(0001), Mg(0001), and Al(111). The distances
of the O$_{2}$ molecule from the three kinds of metal surfaces are fixed at
the same value of 7 \AA , at which charge transfer should not occur in
reality. Also, the O$_{2}$ molecules in the three systems are identically
chosen to be in the HH-Z entrance. The calculated spin-split PDOS for the
O$_{2}$ molecule and the charge-density difference $\Delta\rho(\mathbf{r)}$ in
the three systems are shown in Fig. 8. Clearly, one can see that for O$_{2}%
$/Al(111) and O$_{2}$/Mg(0001), although the distance of the O$_{2}$ molecule
from the surfaces is set to be as large as 7 \AA ~ and charge transfer should
not be presented in real systems, the adiabatic DFT calculations give yet the
contrary results. In these two cases, it reveals in Fig. 8(b) and 8(c) that
the LUMO of the ideal O$_{2}$ molecule (i.e., the spin-down antibonding
$\pi_{p}^{\ast}$ MO) has shifted down to align with the Fermi energy and
becomes partially occupied, accompanying with an observable decreasing of the
molecular spin. The O-O bond length also undergoes a little expansion compared
to the free value of 1.235 \AA . Actually, Fig. 8(b) and 8(c) depict nothing
more than what have been known in previous studies, and it is exactly due to
this unphysical (metal-molecule or intramolecule) charge transfer that
motivates theoretical workers to remedy the calculations by, for example,
nonadiabatically constraining the molecular spins. As a direct result, a
sizeable energy barrier will appear on the corrected PES \cite{Beh2005}. When
the attention is paid to the O$_{2}$/Be(0001) system, however, we find that
such unphysical large-distance charge transfer effect does not happen in the
calculation. In fact, from Fig. 8(a) one can see that at a metal-molecule
distance of 7 \AA , the calculated spin-split molecular PDOS shows no change
with respect to the free case. The LUMO keeps empty and the charge-density
difference is zero everywhere. Subsequently, the molecular bond length and
spin of O$_{2}$ are not influenced at all by the presence of the Be(0001)
surface at a distance of 7 \AA . Combining with Fig. 7, therefore, we conclude
that unlike Al(111) and Mg(0001), the adiabatic DFT calculation presented in
this paper is sufficient to predict an activated-type dissociation process of
O$_{2}$ at Be(0001). \begin{figure}[ptb]
\begin{center}
\includegraphics[width=0.6\linewidth]{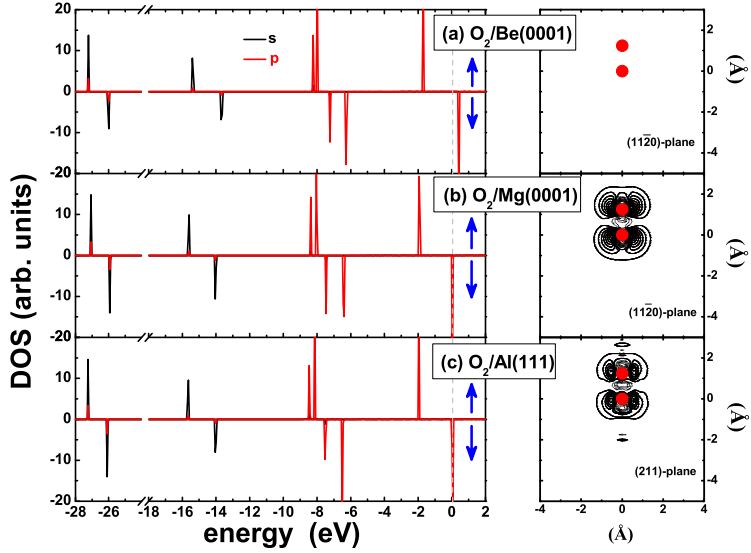}
\end{center}
\caption{(Color online) spin-polarized DOS for the molecular orbits of O$_{2}$
with a fixed vertical distance of 7 \AA ~ from (a) Be(0001), (b) Mg(0001), and
(c) Al(111) surfaces. The O-O bond length in each case is allowed to relax.
The corresponding charge-density differences $\Delta\rho(\mathbf{r})$ are
plotted on the right side with the contour spacing of 0.002/\AA $^{3}$. Solid
and dotted lines denote the accumulated and depleted densities, respectively.}%
\end{figure}

For further illustration of O$_{2}$ dissociation at Be(0001), we plot in Fig.
9 four snapshots for the spin density (the contour spacing is 0.2 $\mu_{B}%
$/\AA $^{3}$) evolving along the dissociation path shown in Fig. 7. The values
of the corresponding heights $h$ and relaxed bond lengths $d$ of the O$_{2}$
molecule are also indicated in the figure. One can see from Fig. 9 that as the
O$_{2}$ molecule approaches the surface along the T-Y channel, the O-O bond
length increases, while the spin magnetic moment decreases and tends to vanish
at $h\mathtt{\sim}$2.3 \AA . Since there has occurred electron transfer from
the substrate to the molecule at this height, thus the disappearance of the
spin moment is due to the donation of the minority spins from the substrate
not due to the spin flip inside the molecule.

\begin{figure}[ptb]
\begin{center}
\includegraphics[width=0.8\linewidth]{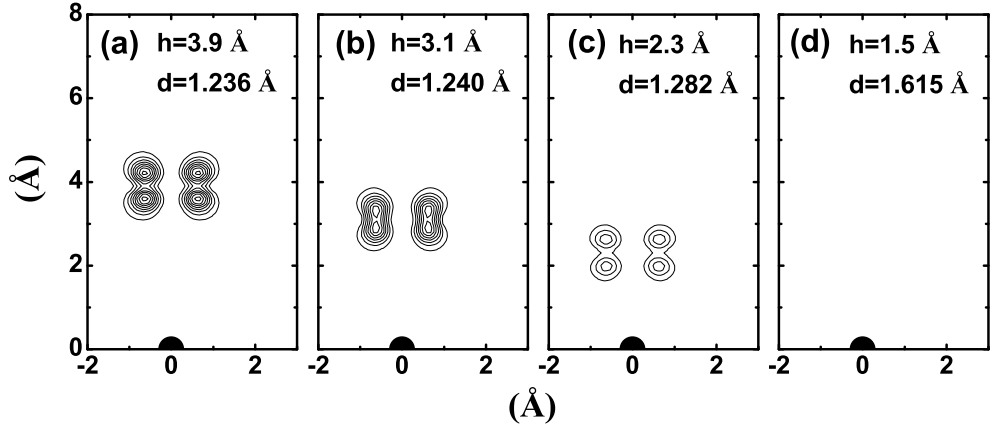}
\end{center}
\caption{Snapshots of the spin quenching process during O$_{2}$ dissociation
at Be(0001) along the T-Y channel. The spacing of the spin-density contours is
0.2 $\mu_{B}$/\AA $^{3}$. The corresponding O$_{2}$ molecular bond length $d$
and distance $h$ from the surface are also indicated. }%
\end{figure}

\begin{figure}[ptb]
\begin{center}
\includegraphics[width=0.8\linewidth]{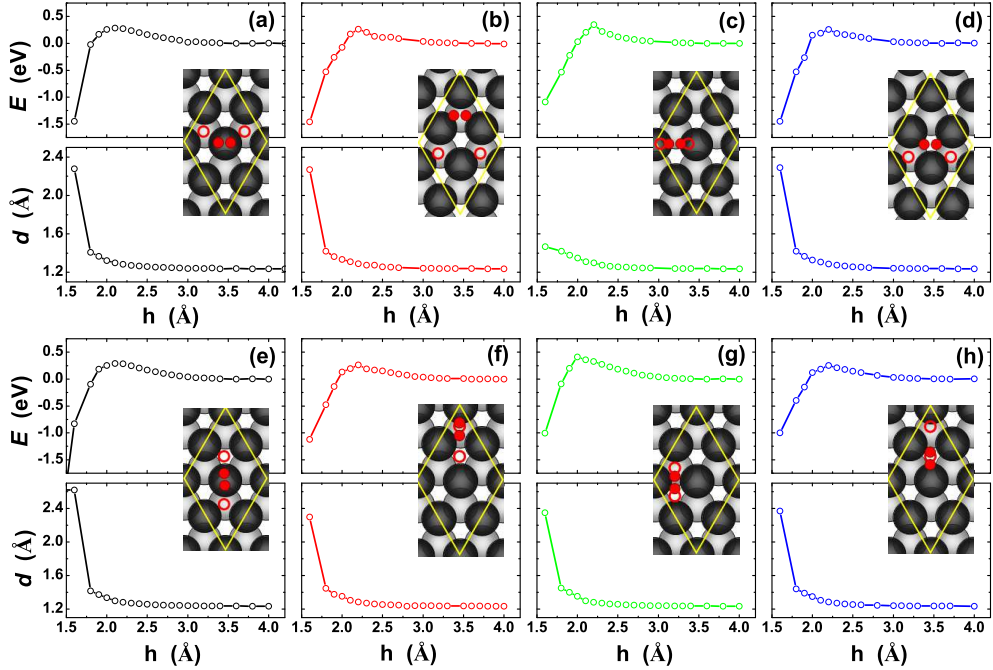}
\end{center}
\caption{One-dimensional cuts of the potential-energy surfaces and the
corresponding O-O bond lengths as functions of the O$_{2}$ distance $h$ from
the surface, for eight different dissociative channels. The inset in each
panel indicates the initial (filled circles) and final (hollow circles) atomic
positions of O$_{2}$.}%
\end{figure}

The adiabatic PES's for O$_{2}$ dissociation at Be(0001) along the other
parallel channels have also been calculated, which are found to have very
similar elbow shapes except for visible differences in the transition states
and dissociation barriers. To be more clear, a detailed comparison of these
PES's is depicted in Fig. 10 by separately plotting the one-dimensional cuts
of PES's and O-O bond length as functions of the O$_{2}$ height $h$ from the
substrate surface. The insets in each panel in Fig. 10 display the initial and
final positions of the two O atoms. Clearly, the dissociative adsorption of
O$_{2}$ along each of these parallel channels is a direct and activated type.
The obtained dissociative adsorption energies, the dissociative energy
barriers, and the geometrical parameters for the corresponding transition
states, including the height of the O$_{2}$ molecule and the O-O bond length,
are concluded in Table II. It is found that the dissociation path with both
the lowest energy barrier and the largest dissociative adsorption energy is
along the T-Y channel. For all the dissociation paths, the O-O bond lengths at
the transition states are similar, ranging from 1.28 to 1.30 \AA , elongated
from the 1.236 \AA ~ in an isolated O$_{2}$ molecule. Note that no
dissociative adsorptions are found for the O$_{2}$ molecules along the
vertical entrances, which thus are not plotted in Fig. 9 and listed in Table
II. This is different from the O$_{2}$/Al(111) system, in which the molecular
dissociation along the vertical channels has been predicted to occur as well
\cite{Honkala}.

\begin{table}[ptb]
\caption{The calculated dissociative adsorption energy ($E_{ad}$),
dissociative energy barrier ($\Delta E$), and geometric parameters (namely,
the O$_{2}$ molecular bond length $d$ and distance $h$ from the surface) of
the transition state, along the eight different dissociation channels.}%
\label{table2}
\begin{tabular}
[c]{ccccc}\hline\hline
Channel & $E_{ad}$ & $\Delta E$ & $h$ (TS) & $d_{\text{O}-\text{O}}$
(TS)\\\hline
T-X & 4.00 & 0.28 & 2.10 & 1.30\\
T-Y & 4.14 & 0.23 & 2.10 & 1.30\\
B-X & 3.99 & 0.26 & 2.20 & 1.29\\
B-Y & 4.10 & 0.26 & 2.20 & 1.29\\
HH-X & 3.81 & 0.35 & 2.20 & 1.30\\
HH-Y & 4.09 & 0.41 & 2.10 & 1.30\\
FH-X & 4.00 & 0.25 & 2.20 & 1.29\\
FH-Y & 4.10 & \ 0.25 & 2.20 & 1.28\\\hline\hline
\end{tabular}
\end{table}

\section{CONCLUSION}

In summary, by performing \textit{ab initio} simulations we have for the first
time systematically investigated the adsorption and dissociation of oxygen
molecules on the Be(0001) surface. We have identified both the physisorbed and
the chemisorbed molecular states, which directly occur along the parallel and
vertical channels, respectively. For the most stable chemisorbed molecular
state (which is along the HH-Z entrance), in particular, we have studied its
electronic and magnetic properties by calculating the charge-density
difference, the spin density, and the PDOS, which clearly show the charge
transfer from the spin-up $\pi^{\ast}$ MO to the substrate followed by the
back donation from the substrate to the spin-down $\pi^{\ast}$ MO, as well as
the distinct covalent weight in the chemical bonding between O$_{2}$ and
Be(0001). This important covalent weight in the molecule-metal bond has been
revealed through the combining fact: (i) The charge is largely accumulated
along the Be-O bond [Fig. 5(a)]; (ii) The $\sigma$, $\sigma^{\ast}$, and $\pi$
MO's are heavily hybridized with the Be $sp$ states (Fig. 6); (iii) The
chemisorbed molecular state in the HH-Z channel is much stable than that in
the FH-Z channel, implying an observable role played by the subsurface Be atom
beneath the O$_{2}$ molecule.

The energy path for the dissociation of O$_{2}$ on Be(0001) surface has been
determined by calculating the adiabatic PES's along various channels, among
which the T-Y channel has been found to be the most stable and favorable for
the dissociative adsorption of O$_{2}$. Remarkably, our results have shown
that the adiabatic dissociation process in the present O$_{2}$/Be(0001) system
is an activated type, with the lowest energy barrier of 0.23 eV in the most
stable T-Y channel. To our knowledge, this is the first time to theoretically
predict a sizeable adiabatic energy barrier during dissociation of O$_{2}$ at
the metal surfaces with simple $sp$ bands. Thus as a final concluding remark,
here we point out that in spite of many previous revealing and specific
studies, an in-depth and primary insight into the common nature of the O$_{2}$
dissociation at various simple $sp$ metal surfaces remains yet to be attainable.

\begin{acknowledgments}
This work was supported by the NSFC under grants No. 10604010 and No.
60776063, and by the National Basic Research Program of China (973 Program)
under grant No. 2009CB929103.
\end{acknowledgments}

\end{document}